\documentclass[reprint,amssymb,amsmath,aip,cha]{revtex4-1}
\usepackage{graphicx}
\usepackage{dcolumn}
\usepackage{bm}
\usepackage[colorlinks=true,linkcolor=blue]{hyperref}%
\expandafter\ifx\csname package@font\endcsname\relax\else
 \expandafter\expandafter
 \expandafter\usepackage
 \expandafter\expandafter
 \expandafter{\csname package@font\endcsname}%
\fi
\usepackage{verbatim}
\usepackage[version=3]{mhchem}
\usepackage{subfigure}

\begin{document}


\title[2D-3D growth transition of ultra-thin Pt films]{Temperature-dependent 2D-3D growth transition of ultra-thin Pt films deposited by PLD}
\thanks{The authors wish to acknowledge financial supported by the Swiss Bundesamt f\"ur Energie (BfE) and Swiss Electric Research (SER) and would like to thank Ulrich M\"uller (Scanning Probe Microscopy User Laboratory, Empa) for his support.}

\author{Henning Galinski}
\email{henning.galinski@mat.ethz.ch}
\homepage{http://www.nonmet.mat.ethz.ch}
\affiliation{Nonmetallic Inorganic Materials, ETH Zurich, Zurich, Switzerland}
\author{Thomas Ryll}
\author{Philipp Reibisch}
\author{Lukas Schlagenhauf}
\author{Iwan Schenker}
\author{Ludwig J. Gauckler}

\date{\today}

\begin{abstract}
During the growth of metal thin films on dielectric substrates at a given deposition temperature $T_d$, the film's morphology is conditioned by the magnitude and asymmetry of up- and downhill diffusion. Any severe change of this mechanism leads to a growth instability, which induces an alteration of the thin film morphology. In order to study this mechanism, ultra-thin Pt films were deposited via pulsed laser deposition (PLD) onto yttria-stabilized-zirconia single crystals at different deposition temperatures. The morphological evolution of Pt thin films has been investigated by means of scanning electron microscopy (SEM), atomic force microscopy (AFM) and standard image analysis techniques. The experimentally obtained morphologies are compared to simulated thin film structures resulting from a two-dimensional kinetic Monte Carlo (KMC) approach.
Two main observations have been made:
i) thin Pt films deposited onto \ce{ZrO2} undergo a growth transition from two-dimensional to three-dimensional growth at $T_d>573$~K. The growth transition and related morphological changes are a function of the deposition temperature. 
ii) A critical cluster size of $i^{*}=4$ in combination with an asymmetric Ehrlich-Schwoebel (ES) barrier favoring the uphill diffusion of atoms allows for a computational reproduction of the experimentally obtained film morphologies.

\end{abstract}

\maketitle

\section{Introduction}
\label{intro}
Pt thin films are commonly used as gate material on metal-insulator-metal (M-I-M) devices~\cite{Rossel1}, auto-catalysts~\cite{Twigg1} and electrodes in micro-solid oxide fuel cells~\cite{Hansen1,Opitz1,Popke1}.
However, thin metal films on dielectric substrates are thermodynamically instable. The instability originates from the conflictive bonding characteristics of metals and insulators~\cite{Finnis1}, which may be expressed by the interfacial energy of metal/dielectric interface, and from the thin film's metastable configuration as a consequence of the chosen deposition method. 
This is especially true for deposition methods like PLD or sputtering, for which the kinetic energy $E_{\text{kin}}$ of the deposited atoms generally exceeds their thermal energy $E_{\text{ther}}$. For the deposited film, this results in a metastable configuration that tends to equilibrate, once subjected to temperature by annealing, joule's heating or radiation. However, if it was possible to keep the deposition rate low, stable thin film structures artificially tailored for specific applications could be achieved in dependence of the deposition temperature $T_d$. This requires the control of the instantaneous deposition flux (initial nuclei density), the average deposition flux (growth rate) and the deposition temperature (diffusion)~\cite{Walton1}. These prerequisites are met for metal growth by pulsed laser deposition (PLD)~\cite{Shen1,Gallivan1}, so called laser-MBE, especially in case of the large ratios between the pulse duration ($\approx 10-25$~ns) and the repetition rate ($\approx 1-100$~Hz).   
This circumstance makes PLD ideal for surface-engineering~\cite{Suo1} and for studying the fundamentals of metal-on-insulator growth.
In order to gain a general understanding of the main principles of metal-on-insulator growth, Pt thin films on single crystalline yttria stabilized zirconia (\ce{ZrO2}) have been chosen as model metal/ceramic systems in the present study. The investigation will focus on how the deposition temperature affects the operative growth mechanism which determines the final thin film morphology. In order to identify the dominating atomistic processes during growth the experiments have been performed alongside with kinetic Monte Carlo simulations.
This paper is structured as follows: Section~\ref{intro} introduced the basic concepts of metal-on-insulator growth and its importance for various applications. Sect.~\ref{sec:2} deals with the experimental framework. In Section~\ref{sec:3}, detailed experimental and simulation results are presented and discussed. The final Sect.~\ref{sec:4} encompasses a summary of the findings and conclusions.

\section{Experimental}
\label{sec:2}

\begin{figure}[h]
	\centering
		\includegraphics[width=0.35\textwidth]{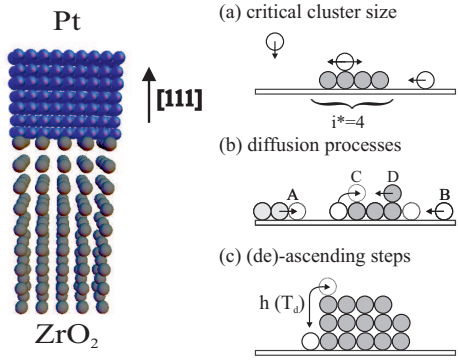}
	\caption[Relevant processes during metal-on-insulator growth]{Schematic illustration of the basic hopping processes used for the KMC model simulating the growth of Pt(111) on \ce{ZrO2}. (a) The growth rate is $1$ atom per $2000$ diffusion steps (exp. $1:4\cdot10^6$) with a critical cluster size $i^{*}=4$ (b) Main processes A: group of atoms split, B: atom diffusion on \ce{ZrO2}, C: atom going up a step, D: atom diffusion on Pt step (c) Temperature-dependent fall or jump height.} 
	\label{fig:growth_modes}
\end{figure}

\subsection{Sample Preparation}
\label{sec:21a}

In order to study the temperature dependence of the growth kinetics of metals on insulators, ultra-thin Pt films were deposited onto single-crystal yttria-stabilised-\ce{ZrO2} (001) via PLD ($F=4.0(2)\text{~J/cm$^2$}$, $p_\text{base}=1\cdot10^{-6}$~mbar, $p_\text{Ar}=2.7\cdot10^{-2}$~mbar). $F$ denotes the laser fluence and $p$ the pressure in the vacuum chamber. The substrates were precleaned using acetone and isopropanol and were thermally smoothed in a muffel furnace for $4$~h at $1673$~K before deposition. The Pt target (purity~$\approx 99.98$~\%) was ablated with a KrF$^+$ excimer laser with a wavelength of $248$~nm, a pulse duration of $25$~ns and a pulse repetition rate of $10$~Hz, leading to a ratio between pulse length and time of $1/4\cdot10^6$.\\
In order to achieve a measurement scheme of sufficient significance, Pt layers with different thicknesses (number of shots $n_\text{s}=25000-250000$) were deposited at various deposition temperatures $T_d$ ranging from $473$ to $1073$~K. It is noteworthy that the Pt/\ce{ZrO2} is suitable as metal-on-insulator growth model system at elevated $T_d$, because it is characterized by a chemically inert interface up to $1273$~K~\cite{Lu1}.\\

\subsection{Kinetic Monte Carlo Simulations}
\label{sec:22}

A conventional two-dimensional kinetic Monte Carlo (KMC) approach has been chosen to reproduce the experimentally observed 2D-3D growth transition. Thereby atoms are deposited on randomly picked sites on a periodic grid with $3000\times30$ atomic positions. The model considers three main kinetic processes: (i) deposition of atoms with a sticking coefficient of $\alpha=0.9$, (ii) the site-specific surface diffusion of adatoms and (iii) the ascending and descending of atoms from steps with different heights.\\
The position-dependence of the diffusion rate is implemented using a variable energy barrier $E_{i\mapsto j}$, similar to the approach of Merrick \textit{et al.} \cite{Merrick1}. For a transition from a site $i$ to a site $j$ the diffusion rate is given by an Arrhenius law and reads as follows
\begin{equation}
D_{i\mapsto j}=\nu_{0,i} \exp\left[-\frac{E_{i\mapsto j}}{k_b T}\right].
\label{eq:diff_rate}
\end{equation}
Thereby the attempt frequency $\nu_{0,i}$ is chosen to be substrate-dependent and is for atoms on Pt $\nu_{0,\text{Pt}}=1.2\cdot10^9$ and $\nu_{0,\text{\ce{ZrO2}}}=6.3\cdot10^9$ for atoms on the \ce{ZrO2} substrate respectively. The barrier height $E_{\text{Pt}\mapsto \text{Pt}}=0.8$~eV for Pt-Pt self-diffusion has been adapted from field-ion microscopy results~\cite{Bassett1}, whereas the barrier height for the Pt diffusion on \ce{ZrO2} has been set to $E_{\text{\ce{ZrO2}}\mapsto \text{\ce{ZrO2}}}=1.0$~eV.\\
The dissociation of unstable clusters of atoms was taken into account by the introduction of a critical cluster size i* = 4. The energy barrier, which corresponds to the enthalpy of the adatom formation, scales with the cluster size accounting for the increased nearest neighbor interaction, see Fig~\ref{fig:growth_modes}~(a).
\begin{figure*}
\begin{center}
 \subfigure[~$T_d=473$~K]{\label{fig:growth-a}\includegraphics[scale=0.33]{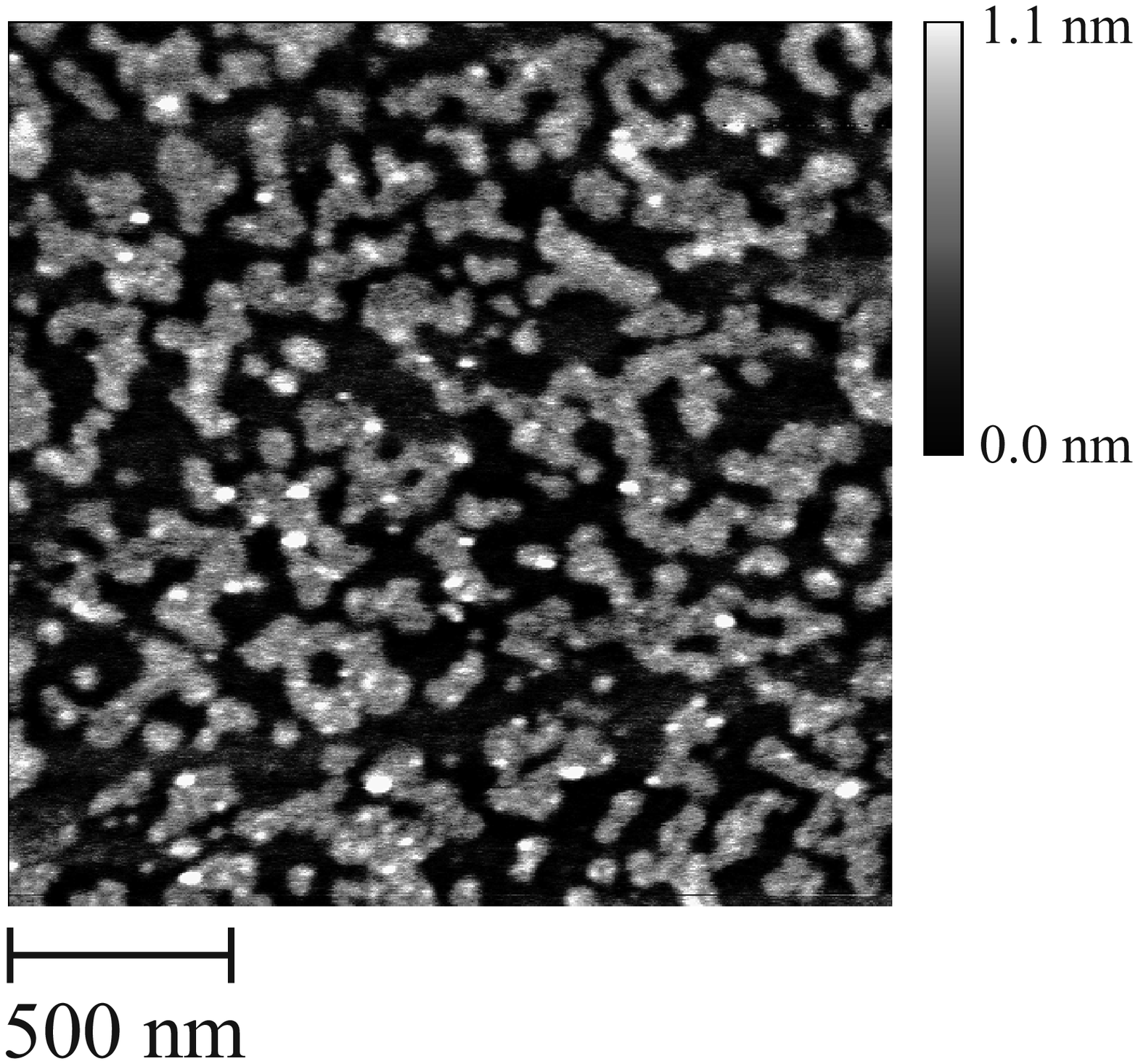}}
 \subfigure[~$T_d=773$~K]{\label{fig:growth-b}\includegraphics[scale=0.33]{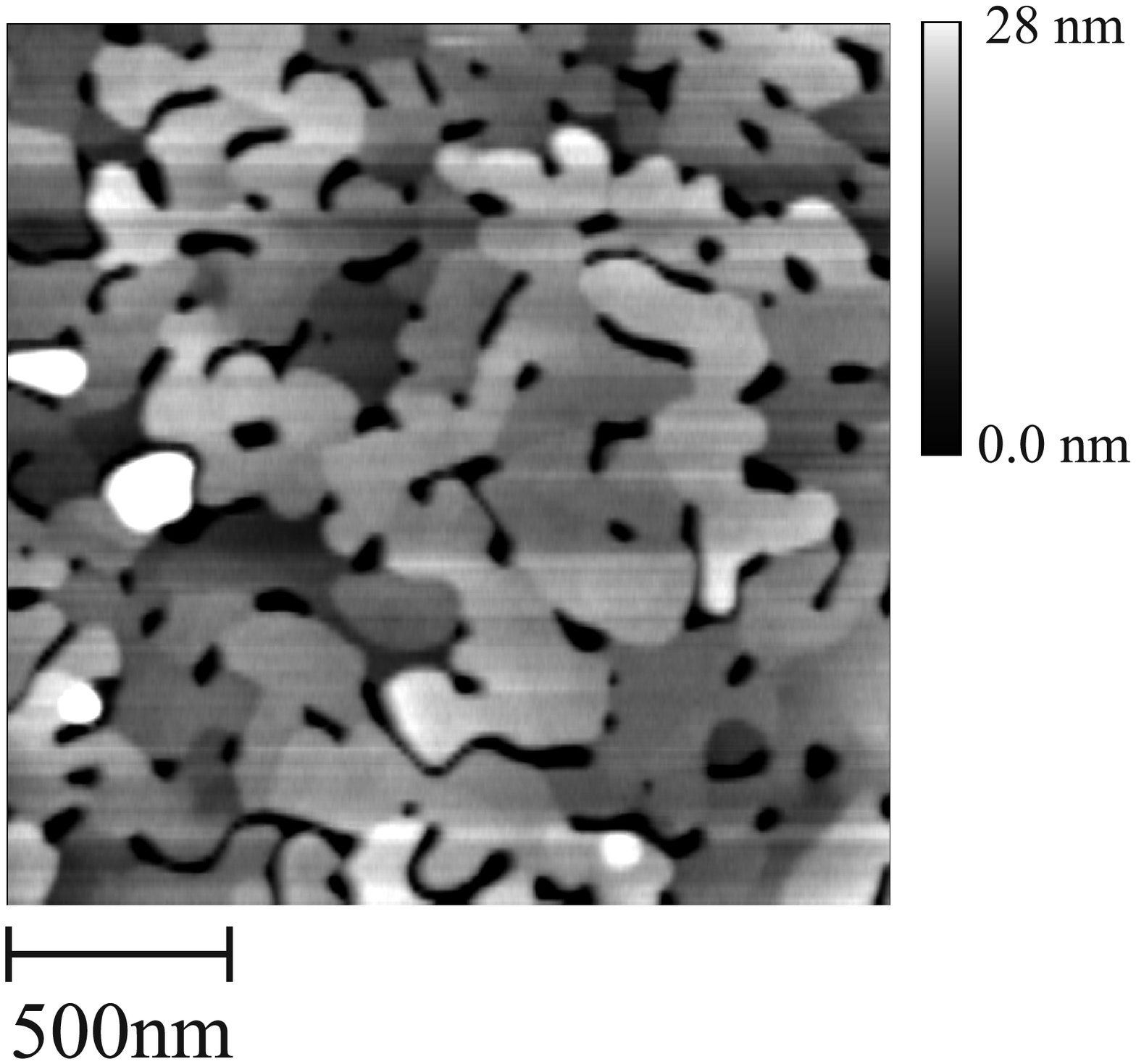}}
 \\
 \subfigure[~$T_d=873$~K]{\label{fig:growth-d}\includegraphics[scale=0.33]{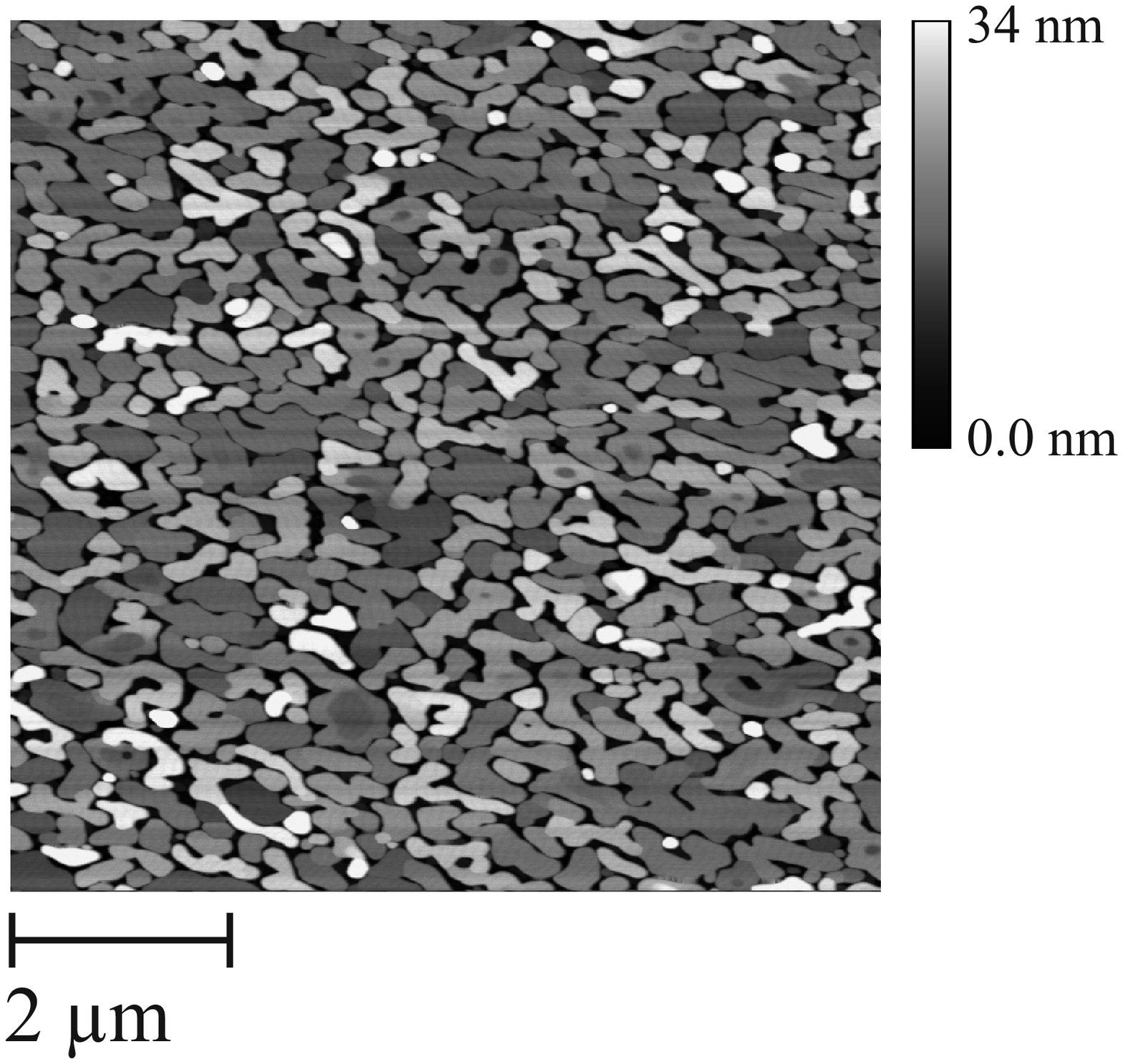}}
 \subfigure[~$T_d=1073$~K]{\label{fig:growth-c}\includegraphics[scale=0.33]{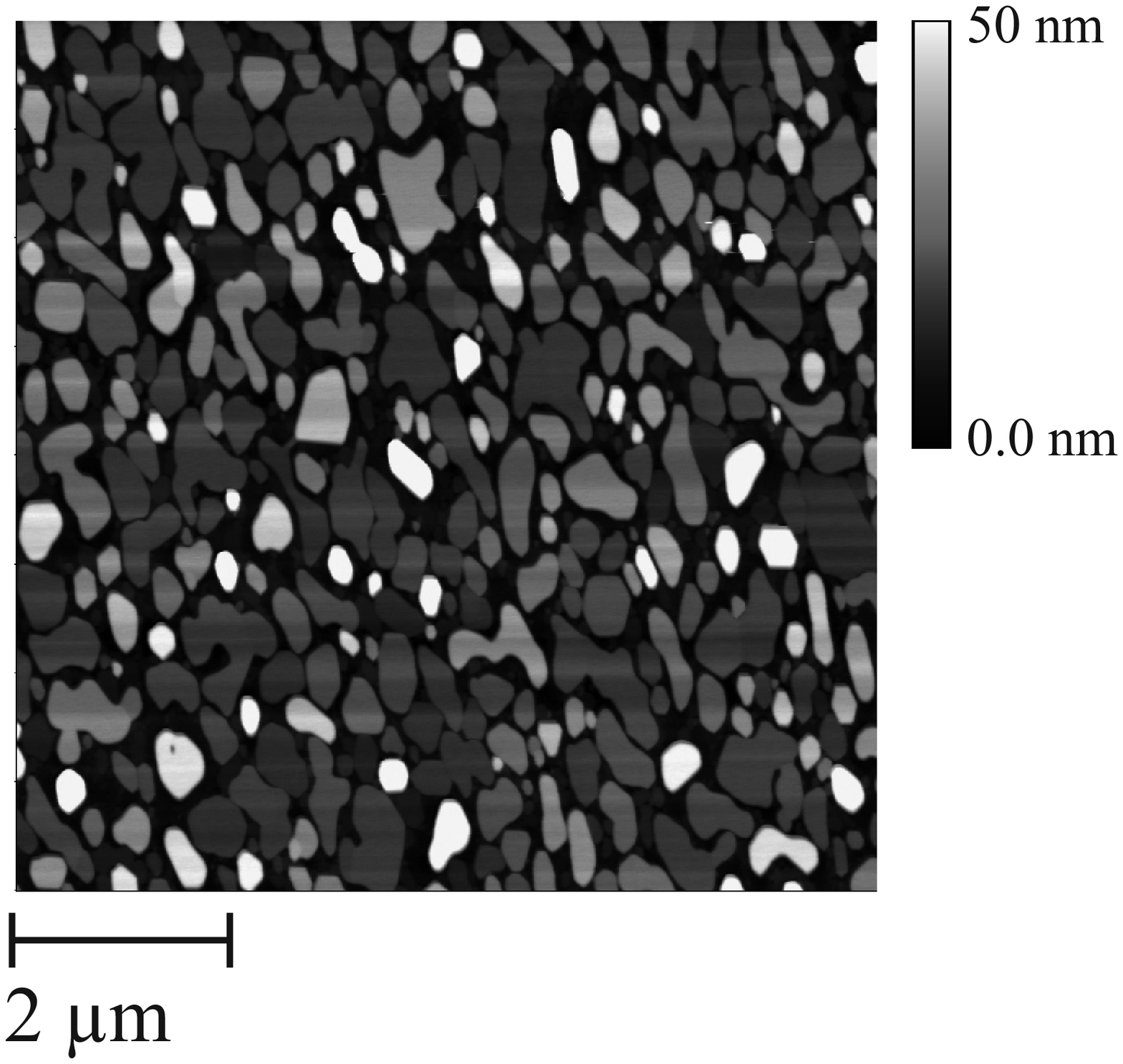}}
\end{center}
\caption[Growth regimes of Pt on \ce{ZrO2}]{Observed growth modes of Pt thin films deposited at different $T_d$ captured by AFM. \subref{fig:growth-a} 2D \textit{layer-by-layer} growth mode with a film thickness $h=16$~nm (determined by RBS). \subref{fig:growth-b} and \subref{fig:growth-d} 3D \textit{kinetically frozen island growth} characterized by a percolating network of elongated islands. \subref{fig:growth-c} 3D \textit{island growth} mode with isolated islands having a mean perimeter of $120(50)$~nm.}
\label{fig:3}       
\end{figure*}
Equivalent to the dissociation of unstable atom clusters, an Ehrlich-Schwoebel (ES) barrier is implemented in the KMC-model that accounts for an asymmetric probability of ascending and descending steps. A detailed overview of the different energy barriers chosen for the different jump processes is given in Tab.~\ref{tab:barriers}.\\
In addition to the ES barrier and in analogy to Leal \textit{et al.}~\cite{Leal1}, the step height $h_{\text{step}}\uparrow~=-14.2+0.03\cdot T_d$, that an atom can overcome by uphill diffusion scales linearly with deposition temperature in the interval $T_{d}=[473,1073]$~K, see Fig~\ref{fig:growth_modes}~(c). However the step height an atom can fall down is kept constant $h_{\text{step}}\downarrow~=2$ for all $T_d>473$~K. The same is true for internal jumps on Pt clusters.

\begin{table}
\caption{Chosen barrier heights used for the various jump processes}
\label{tab:barriers}       
\begin{ruledtabular}
\begin{tabular}{ccc}
process & $E_\text{a}$ on ZrO$_2$ & $E_\text{a}$ on Pt \\
\hline
single atom & 1.0$\,\cdot E_\text{a,ZrO$_2$}$ & 1.0$\,\cdot E_\text{a,Pt}$ \\ 
split of pair & 1.5$\,\cdot E_\text{a,ZrO$_2$}$ & 1.5$\,\cdot E_\text{a,Pt}$ \\ 
split of triple & 2.0$\,\cdot E_\text{a,ZrO$_2$}$ & 2.0$\,\cdot E_\text{a,Pt}$ \\ 
split of group of four & 2.5$\,\cdot E_\text{a,ZrO$_2$}$ & 2.5$\,\cdot E_\text{a,Pt}$ \\ 
going $\uparrow$ one step & 1.5$\,\cdot E_\text{a,ZrO$_2$}$ & 1.5$\,\cdot E_\text{a,Pt}$ \\ 
going $\uparrow$ a group & 2.0$\,\cdot E_\text{a,ZrO$_2$}$ & --- \\ 
single atom going $\downarrow$ & --- & 2.0$\,\cdot E_\text{a,Pt}$ \\ 
atom at a pair going $\downarrow$ & --- & 2.5$\,\cdot E_\text{a,Pt}$ \\ 
atom at a triple going $\downarrow$ & --- & 3.0$\,\cdot E_\text{a,Pt}$ \\
atom at a group of four going $\downarrow$ & --- & 3.5$\,\cdot E_\text{a,Pt}$ \\
\end{tabular}
\end{ruledtabular}
\end{table}

\subsection{Sample Characterization}
\label{sec:21b}

The morphology of the samples was analyzed via high resolution AFM, using a Mobile S (Nanosurf) and a Topometrix 2000, and standard scanning electron microscopy (SEM). The coverage, the distribution of nucleation centers and the island distribution were calculated from SEM or AFM images using standard particle analysis algorithms. While the film thickness has been quantified by Rutherford backscattering spectrometry (RBS), the orientation of the grown films has been determined using grazing incidence X-ray diffraction (GIXD).\\
The simulated profiles and structures were compared to the films prepared by PLD using common statistical tools for surface description, i.e., the roughness and the height-height correlation function (HHCF). 
\begin{eqnarray}
	R_a &=& \frac{1}{N}\sum_{i=1}^N\Big(z_i-\langle z \rangle\Big) \label{Ra} \\
	R_{rms} &=& \sqrt{\frac{1}{N}\sum_{i=1}^N\Big(z_i-\langle z \rangle\Big)^2} \label{Rms}
\end{eqnarray}
For each output file of the simulation, the average roughness $R_a$ and root mean square roughness $R_{rms}$ were calculated according to Eq.~(\ref{Ra}) and Eq.~(\ref{Rms}), as well as the HHCF, 
\begin{eqnarray}
	hhcf_{exp}(m\cdot d) &=& \sqrt{\frac{1}{N-m}\sum_{i=1}^{N-m}\Big(z_{i+m}-z_i\Big)^2} \label{hhcf} \\
	hhcf_{gauss}(\tau) &=& 2\sigma^2\cdot\left(1-\exp\Big(-\frac{\tau^2}{L_c^2}\Big)\right). \label{fit_hhcf}
\end{eqnarray}
Thereby, $d = x_{i+1} - x_i$ is the distance between neighboring points in units of the grid. The determined HHCF is compared by a least square fitting to a Gaussian-HHCF (Eq.~\ref{fit_hhcf}), which is characterized by the root mean deviation of heights $\sigma$ and the correlation length $L_c$. 

\section{Results and Discussion}
\label{sec:3}

\subsection{Nucleation and Growth}
\label{sec:31}

Pt thin films were grown on yttria-stabilized-\ce{ZrO2} single crystals at a pulse frequency of $f_{\text{p}}=10$~Hz over a range of different deposition temperatures, $T_{d}=\left[473,1073\right]$~K and deposition times of $25000 - 250000$ shots. XRD measurements have shown that the Pt thin film preserves a $(111)_{fcc}//(001)_{sc}$ orientation relation to the \ce{ZrO2} substrate for all $T_d$. The overall growth rate is not affected by $T_d$ and is $1.1(5)$ ML (monolayers) per $1000$ laser pulses or $0.19(8)$~nm/min.
For all deposited Pt thin films, the morphological transition as a function of the deposition temperature $T_d$ has been analyzed \textit{ex situ} via AFM and SEM. 

\subsubsection{Temperature domain}
\begin{figure}
 \includegraphics[width=0.4\textwidth]{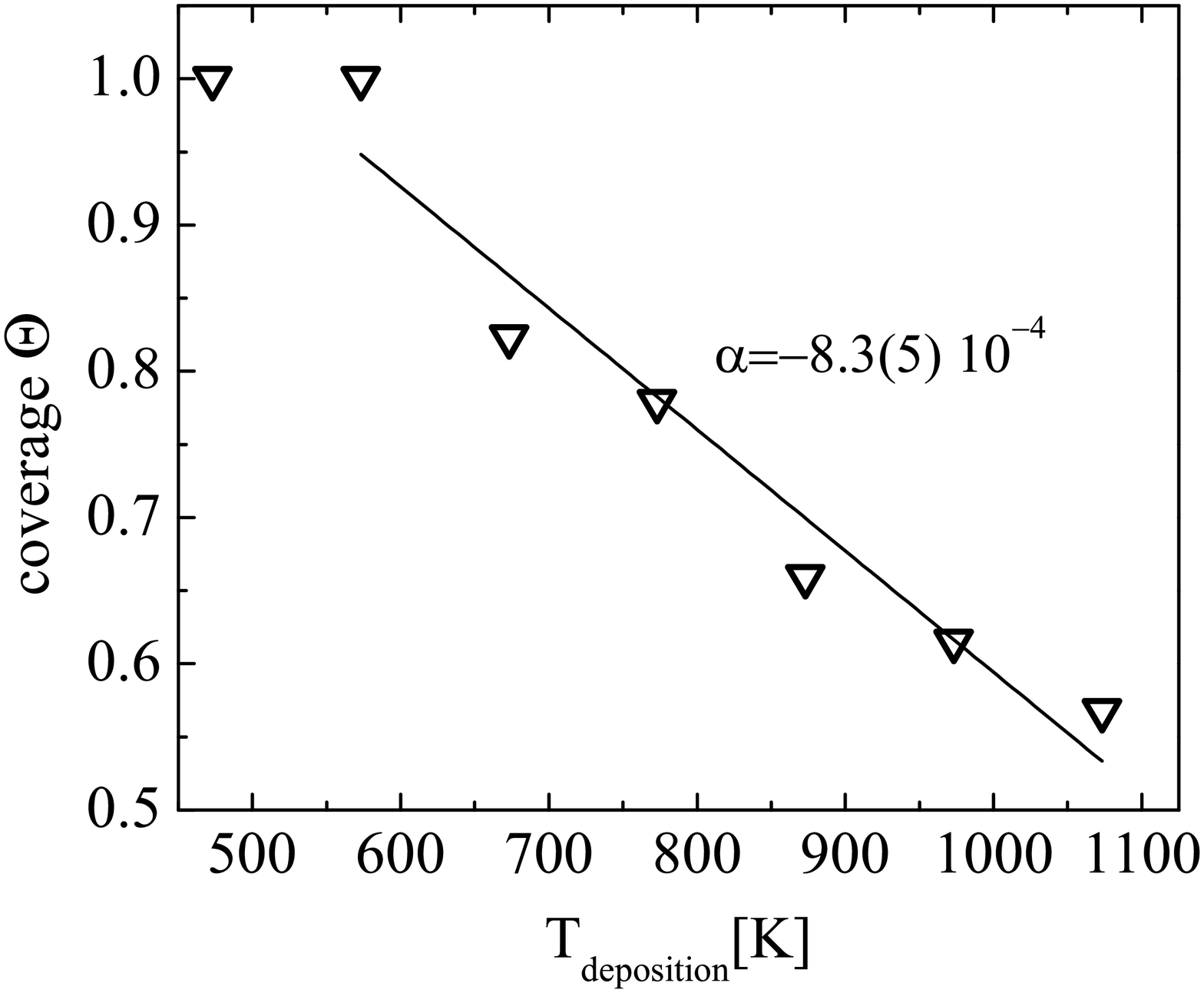}
\caption[Temperature-dependent surface coverage $\Theta$]{Surface coverage $\Theta$ as function of the deposition temperature $T_d$. The coverage decreases linearly for $T_d>573$~K with $\alpha=8.3(5)\cdot 10^{-4}$~K$^{-1}$.}
\label{fig:5}       
\end{figure}
In Fig.~\ref{fig:3}, the experimentally observed three main growth regimes of Pt on \ce{ZrO2} is shown. For sufficiently low deposition temperatures $T_d<473$~K, the thin film grows layer-by-layer, see Fig.~\ref{fig:growth-a}. The film thickness $h$, determined via RBS, is $16$~nm and exceeds significantly the measured height profile of roughly $4$~ML in Fig.~\ref{fig:growth-a}. Therefore, the observed islands can be associated to a yet uncompleted layer of the thin film.\\
Choosing $T_d>573$~K, the film changes its growth regime from a 2D to a 3D growth and undergoes a continuous-discontinuous transition. This is indicated by elongated clusters of grains and holes that are formed (Fig.~\ref{fig:growth-b}) which breaks the symmetry of the film, see Fig.~\ref{fig:growth-d}. Increasing $T_{d}$ further results in an increase in the mean island size and, as a consequence, in a linearly decreasing coverage, see Fig.~\ref{fig:5}. In the last stage, the discontinuous film loses its percolation and an ordered array of faceted particles is formed as shown in Fig.~\ref{fig:growth-d}.\\
In the early stages of film growth, i.e. for $T_{d}\le 573$~K, the surface diffusion and surface self-diffusion of Pt is suppressed. This results in a decreased critical diffusion length $l_c$ (the mean distance between an adatom and a pre-existing nucleation site) and an enhanced density of nucleation sites due to the decreased mobility. It is noteworthy, that the coalescences' kinetics of different stable islands is decelerated as the capillary forces depend on the surface diffusion $D_s$ too~\cite{Evans1}. Therefore, the found layer-by-layer growth has to be understood as a result of kinetic processes rather than thermodynamic relaxation. Based on the assumption that the density of nucleation clusters $\rho_\text{nuc}$ is enhanced and the cluster coalescence is negligibly slow, the probability that atoms are deposited beyond the edge of steps is increased. In combination with preferential downhill funneling~\cite{Evans1,Yu1} at these step edges, this leads to a layer-by-layer growth and a smooth film topography as shown in Fig.~\ref{fig:growth-a}.\\ 
Once the substrate temperature is shifted to higher temperatures, $T_{d}>573$~K, the surface diffusion is increased and the capillary forces gain more impact on the morphological evolution of the thin film. This onset of surface diffusion is featured in the observed 2D-3D growth transition, which can be seen in Fig.~\ref{fig:growth-a} and Fig.~\ref{fig:growth-b}. In the early stage of 3D growth, the formed film morphology can be understood as result of two competing mechanisms: the radial expansion of single clusters and the coalescences of two or more impinging clusters~\cite{Jeffers1}. If the expansion kinetics of those clusters is much faster than the coalescence, the growth is \textit{kinetically frozen}~\cite{Aziz1} and pores form within a percolating network of elongated grains, see Fig.~\ref{fig:growth-b}. With increasing substrate temperature $T_d>773$~K, the porosity and the island size, with a mean perimeter of $d_{island}=120(50)$~nm, increase (Fig.~\ref{fig:growth-d}) until the percolation of the thin film is lost (Fig.~\ref{fig:growth-c}). These findings are in good accordance with the morphological evolution of epitaxial growth of Ag on Mica for different deposition temperatures~\cite{Baski1}.
The present morphologies and especially the decreasing coverage with increasing $T_d$ can well be explained under the assumption that adatoms can jump from the substrate on pre-existing clusters. In a first approximation, the thermal activation of the jump height is assumed to scale linearly with the film coverage (Fig.~\ref{fig:5}).
It is noteworthy that the present morphological evolution at high $T_d$ is presumably due to the same capillary forces, that cause the continuous-discontinuous transition of Pt on \ce{ZrO2} during thin film agglomeration~\cite{Galinski1}.\\

\subsubsection{Time domain}
\begin{figure*}
\begin{center}
 \subfigure[~25000 shots, $T_d=773$~K]{\label{fig:kinks-a}\includegraphics[width=0.30\textwidth]{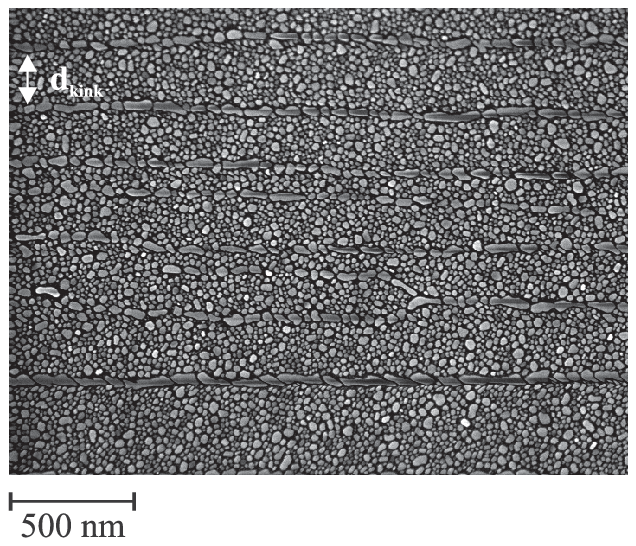}}
 \subfigure[~50000 shots, $T_d=773$~K]{\label{fig:kinks-b}\includegraphics[width=0.30\textwidth]{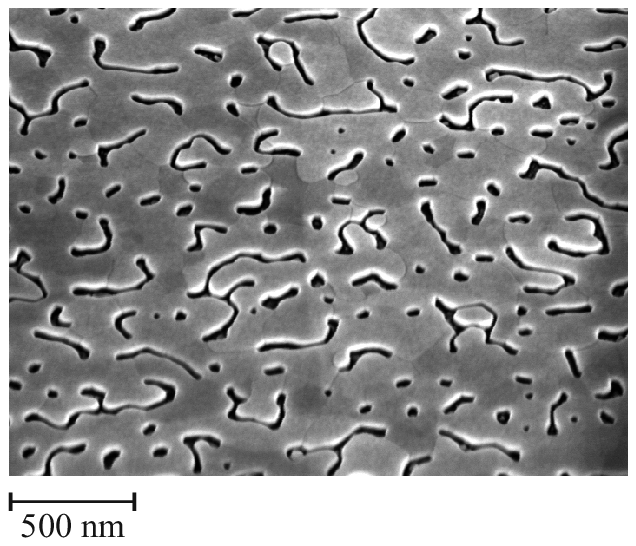}}
 \subfigure[~250000 shots, $T_d=773$~K]{\label{fig:kinks-c}\includegraphics[width=0.30\textwidth]{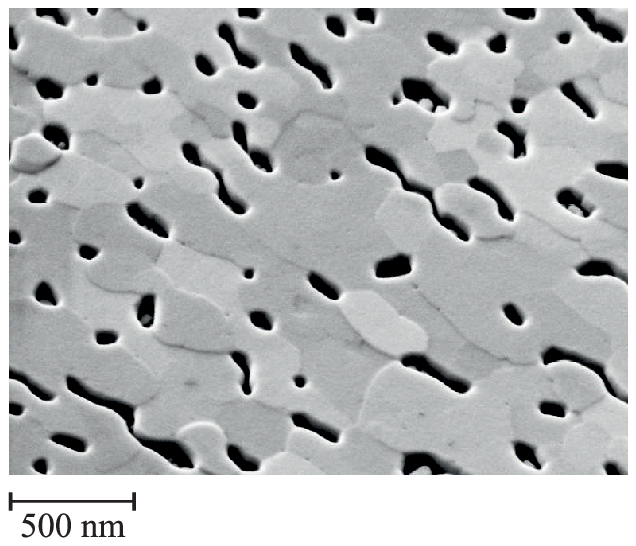}}
\end{center}
\caption[Nucleation and growth of Pt on \ce{ZrO2}]{SEM images of Pt film growth at $T_d=773$~K for different times (number of shots). \subref{fig:kinks-a} Nucleation of islands on a vicinal terraced substrate with enhanced nucleation at the terrace steps. The mean terrace width is $d_{\text{kink}}=245(8)$~nm. \subref{fig:kinks-b} and \subref{fig:kinks-c} 3D \textit{kinetically frozen island growth}, characterized by a percolating network of elongated islands.}
\label{fig:2}       
\end{figure*}
In order to compare the obtained growth morphologies with findings on metal-on-insulator growth via PLD in literature, a set of films were deposited at $T_d=773$~K for three different deposition times. The resulting morphology evolution as function of deposition time, i.e. number of shots, is depicted in Fig.~\ref{fig:2}. 
In Fig.~\ref{fig:kinks-a}, the initial stage of film growth is shown. Isolated Pt nuclei with a mean next-neighbor distance of $l_{eff}=32(1)$~nm have been formed on the terraces and along the step edges of the vicinal \ce{ZrO2} substrate. The density of nuclei per unit area was determined to be $\rho_{\text{nuc}}=609$~$\mu$m$^{-2}$. The enhanced growth at the step edges is in good accordance with results from literature on the epitaxial growth of Pt $(111)$ on $(0001)$ sapphire~\cite{Farrow1,Ramanathan1}. In these cases, it has been found that the sapphire surface steps act as nucleation sites and cause rotational twinning in the Pt-film.
With increasing deposition time, the nuclei grow and coalesce with each other, see Fig.~\ref{fig:kinks-b}. Because of the different velocities of the cluster growth and the coalescences of clusters the resulting film is disrupted by holes, i.e.~the growth is kinetically frozen~\cite{Jeffers1}. Further deposition as shown in Fig.~\ref{fig:kinks-c}, increases both the feature size of impinging grains and the hole size. In addition, new nucleation clusters form inside the cavities of the disrupted film. The found growth morphologies and their progression with increasing deposition time are in good agreement with the experimental findings on metal-on-insulator growth of Ag on Mica~\cite{Aziz1,Aziz2,Aziz3}.      

\subsection{Experiment vs. Simulation}
\label{sec:32}

\begin{figure}
 \includegraphics[width=0.4\textwidth]{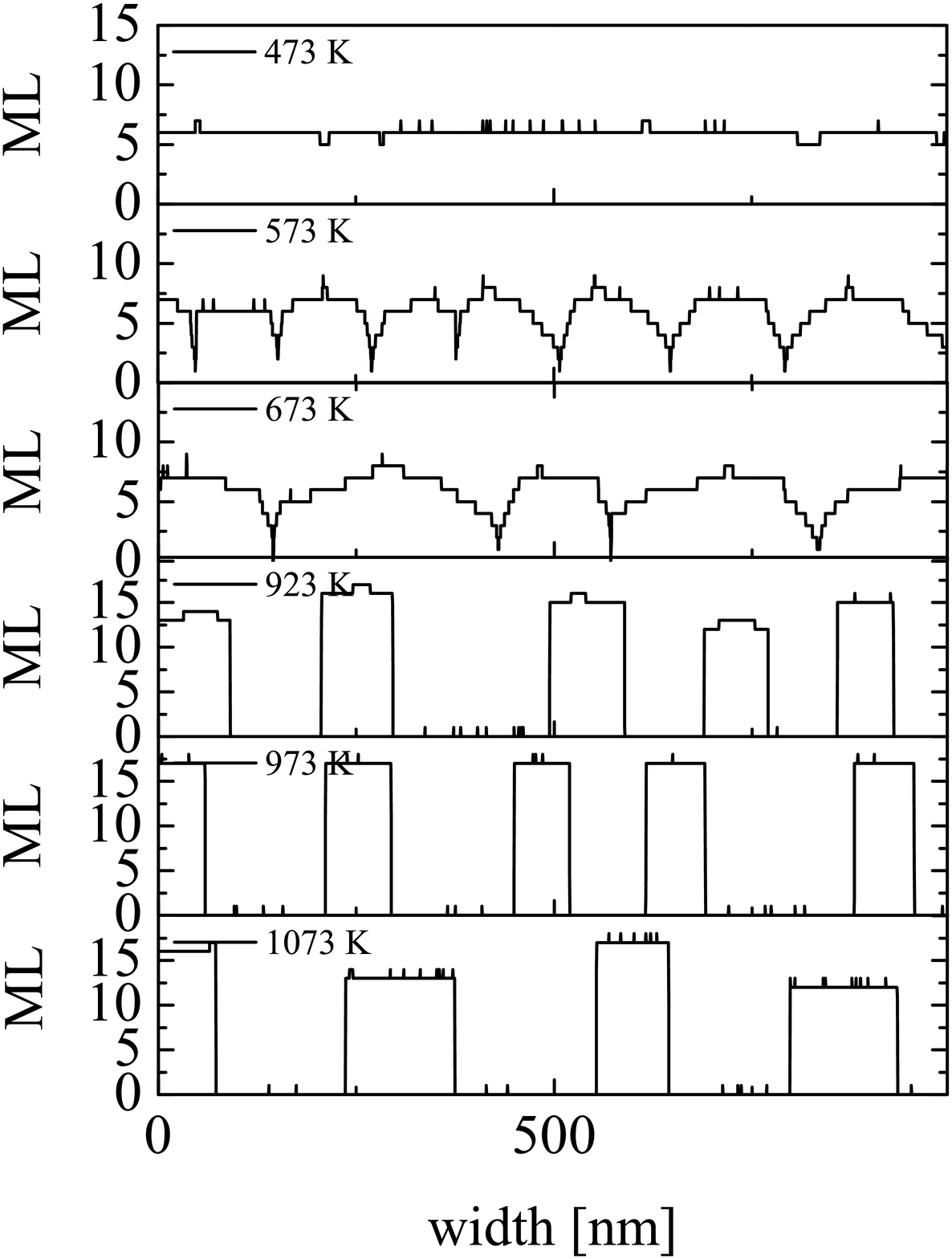}
\caption[Film morphologies obtained by KMC]{Surface morphologies obtained from KMC simulations after $20000$ growth steps for 6 different deposition temperatures $T_d$. The growth conditions as well as the obtained surface morphologies scale with altered substrate temperature $T_d$.}
\label{fig:4}       
\end{figure}

In order to substantiate the previous findings, the significance of an increasing uphill diffusion with increasing deposition temperature $T_d$ regarding the morphological progression of Pt on \ce{ZrO2} has to be confirmed. In Fig.~\ref{fig:4}, the morphological regimes during thin film growth, simulated by the two-dimensional KMC model, are shown. The chosen approach reproduces the experimentally observed transition from \textit{layer-by-layer} growth to 3D \textit{mound formation} at $T_d\approx 573$~K. The continuous-discontinuous transition of the thin film for $T_d>673$~K has been reproduced as well. Thus, it can be concluded that the experimentally observed growth transitions can be treated in terms of a thermally activated step height dependent ES barrier, whereby the jump height of a diffusing adatom increases linearly with $T_d$.
\begin{figure}
\begin{center}
 \subfigure[]{\label{fig:vs-a}\includegraphics[width=0.4\textwidth]{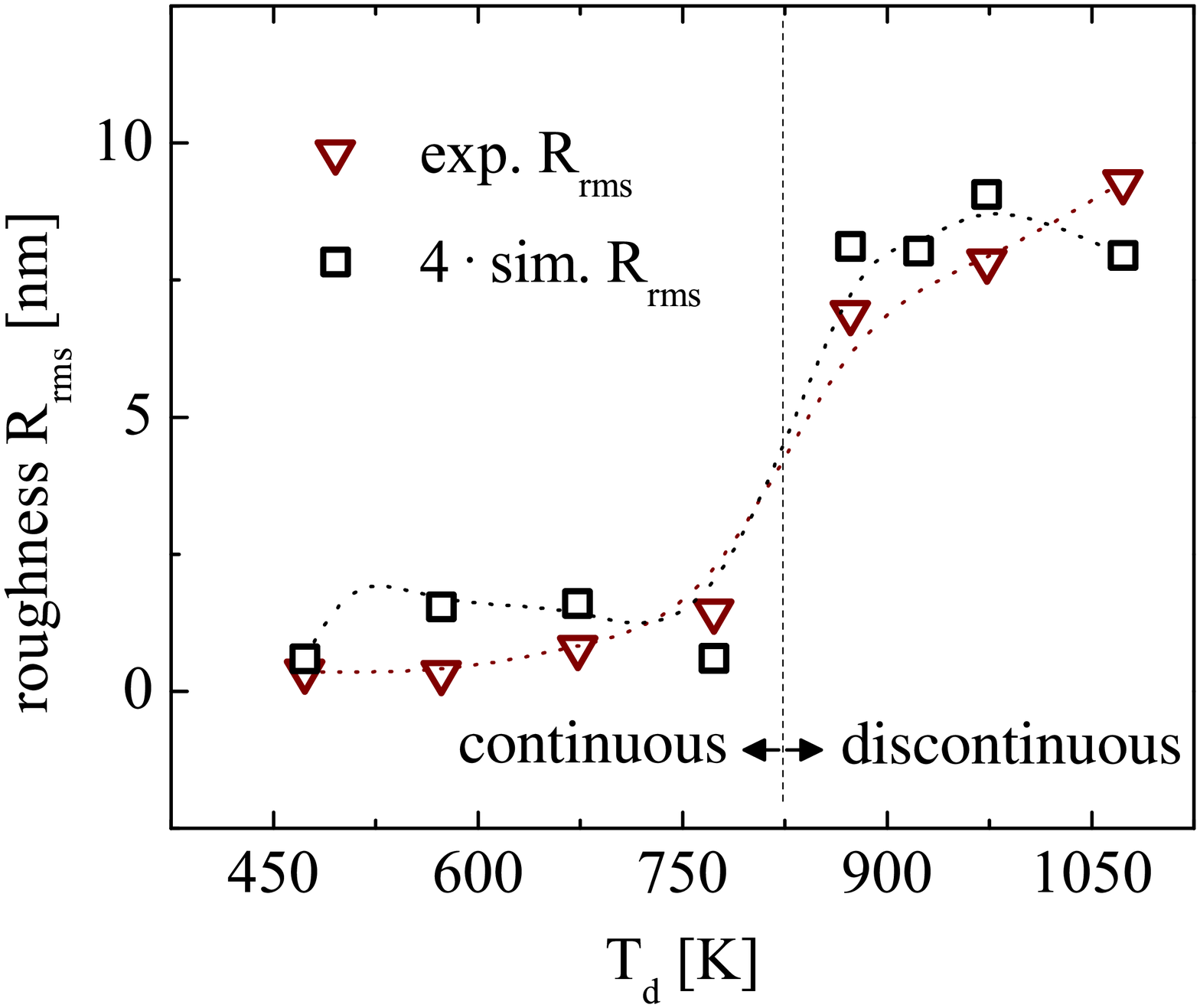}}\\
 \subfigure[]{\label{fig:vs-b}\includegraphics[width=0.4\textwidth]{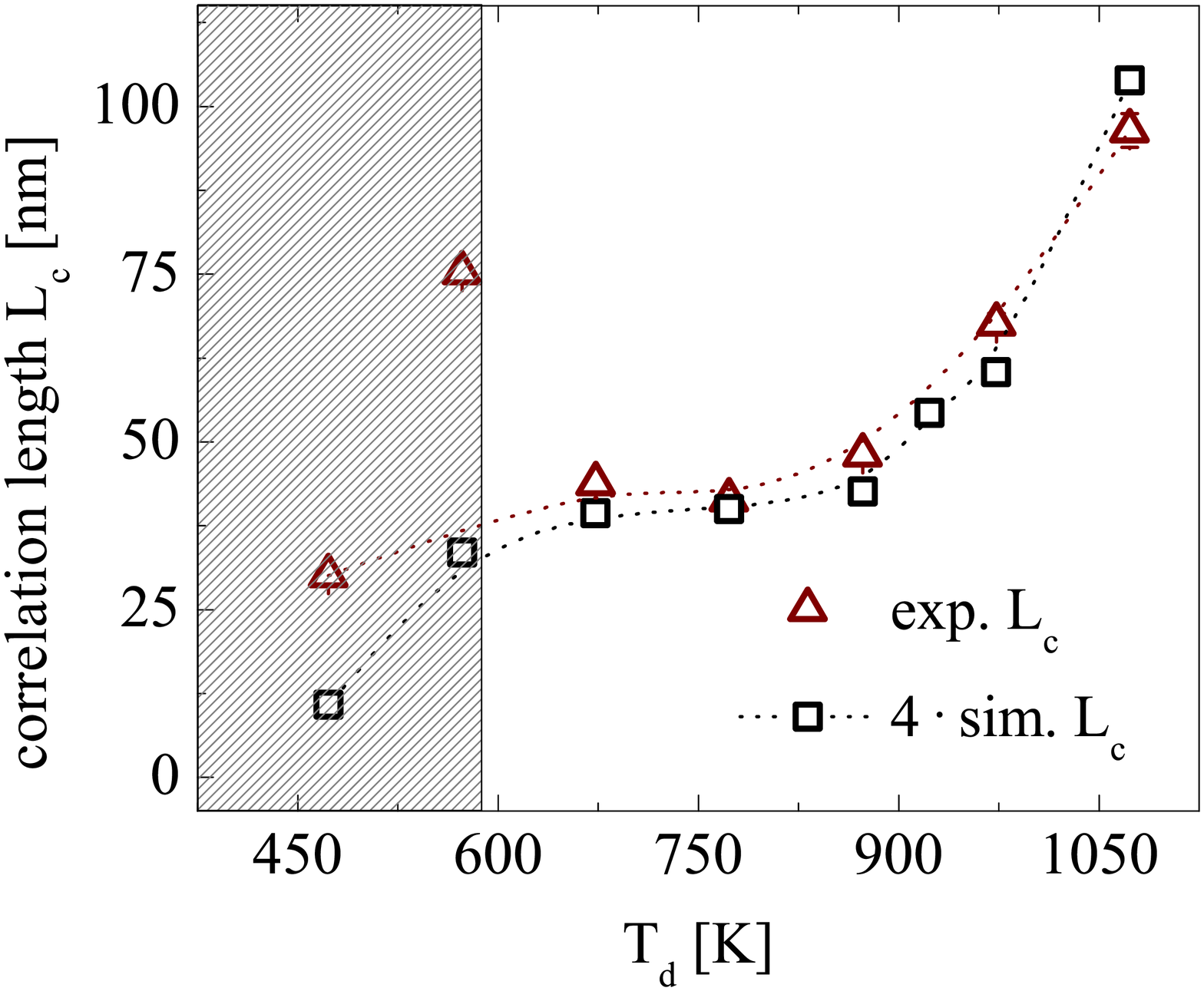}}
\end{center}
\caption[Comparison of film morphologies (experiment vs. simulation)]{\subref{fig:vs-a} Plot of the root mean square roughness R$_{\text{rms}}$ vs. $T_d$ for the experimentally grown and the simulated Pt thin film morphologies. \subref{fig:vs-b} Plot of the correlation length $L_c$ as function of $T_d$, comparing the experimental with KMC simulation results. The gray shaded area denotes the 2D growth regime}
\label{fig:6}       
\end{figure}
In order to quantitatively validate the morphological resemblance between the experimental and simulation results, the HHCF was chosen measuring the scaling properties and roughness for all obtained thin films. The HHCF captures two main surface characteristics, the high frequency fluctuations perpendicular to the surface represented by the root mean deviation of heights $\sigma$ and the correlation length $L_c$. The latter denotes the critical distance, within characteristics of the surface can be expected to be correlated~\cite{Palasantzas1,Constantoudis1}.   
The measured RMS-roughness for the experimental and the simulated surfaces is shown in Fig.~\ref{fig:vs-a}. The step-like profile coincides with the observed continuous-discontinuous transition driven by the enhanced capillary forces. The RMS-roughness for the experimental and the simulated surfaces follows the same trend. 
The observed difference by a factor of four between the experimental data (deposition on a 2D surface) and simulated data (deposition on a 1D line) is attributed to dimensional effects. A similar conclusion can be drawn regarding the evolution of $L_c$ as a function of the deposition temperature $T_d$: $L_c$ is in the order of the film's grainsize for low $T_d$ and of the island size for high $T_d$. The surface characteristics of both experimental and simulation results are in good accordance, so that it can be concluded that the main growth mechanisms have been found for the present metal-insulator couple.
\section{Conclusion}
\label{sec:4}

In essence, it has been shown that Pt thin films deposited by PLD undergo a 2D-3D growth transition as function of the deposition temperature $T_d$. The growth kinetics and morphological evolution depend strongly on the onset of the surface self-diffusion and its related capillary forces.
Three main growth regimes have been identified: i) \textit{layer-by-layer} growth governed by downward funneling due to the absence of capillary forces, ii) \textit{kinetically frozen island} growth, which is attributed to the onset of surface diffusion that competes with nuclei growth and coalescence and iii) pure 3D island growth, which is dominated by uphill diffusion due to capillary forces. The choice of the appropriate deposition temperature thus allows for an application-specific tailoring of the thin film's structure and hence its properties. In particular, these morphological transitions are in excellent agreement with the results of temperature-dependent growth experiments of Ag on Mica by Baski and Fuchs~\cite{Baski1}.
Additionally, the temperature-dependent growth kinetics and morphological evolution of ultra-thin Pt films have been investigated numerically using Kinetic Monte Carlo simulations. The simulations were performed using a critical cluster size of $i^{*}=4$ and an asymmetric, temperature-dependent ES barrier facilitating the upward diffusion of Pt atoms on pre-existing terraces. This allowed for a qualitative reproduction of the experimentally determined surface roughness and height height correlation length, which confirms the pre-dominant impact of capillary forces on the thin film growth at elevated deposition temperatures. 
It has been shown that a detailed knowledge of the interacting physical growth mechanism is essential to control the engineering of thin films at elevated deposition temperature. Within this manuscript the capillary forces, which also control solid state dewetting phenomena~\cite{Galinski1}, have been identified as the most crucial.


\bibliography{pld_arxiv}

\end{document}